\documentclass[12pt, preprint]{aastex}
\usepackage{graphicx}

\shortauthors{B. Paul et al.}
\shorttitle{GX~1+4}

\begin{document} 

\title{High resolution X-ray spectrum of the accreting binary X-ray
pulsar GX~1+4}

\author{B. Paul\altaffilmark{1,2}, 
T. Dotani\altaffilmark{1}, F. Nagase\altaffilmark{1},
U. Mukherjee\altaffilmark{2}, and S. Naik\altaffilmark{3}}

\altaffiltext{1}{The Institute of Space and Astronautical Science,
3-1-1 Yoshinodai, Sagamihara, Kanagawa, 229-8510, Japan.
nagase@astro.isas.jaxa.jp, dotani@astro.isas.jaxa.jp}

\altaffiltext{2}{Department of Astronomy and Astrophysics,
Tata Institute of Fundamental Research, Homi Bhabha road, Mumbai,
400\,005, India. bpaul@tifr.res.in, uddipan@tifr.res.in}

\altaffiltext{3}{Department of Physics, University College Cork, Cork,
Ireland, sachi@ucc.ie}


\begin{abstract}

We present here high resolution X-ray spectrum of the accreting binary
X-ray pulsar GX~1+4 obtained with the High Energy Transmission Grating (HETG)
instrument of the Chandra X-ray Observatory. This was supplemented by a
simultaneous observation with the Proportional Counter Array (PCA) of the
Rossi X-ray Timing Explorer (RXTE). During this observation, the source was
in a somewhat low intensity state and the pulse profile with both Chandra
and RXTE shows a narrow dip, characteristic of GX~1+4 in medium and low
intensity states. The continuum X-ray spectrum obtained with the HETG and PCA 
can be fitted well with a high energy cutoff power-law model with line of
sight absorption. Interestingly, we find that this low state is accompanied
by a relatively small absorption column density. A 6.4 keV narrow emission
line with an equivalent width of 70 eV is clearly detected in the HETG
spectrum. The fluorescence iron line, or at least part of it is produced
in the neutral or lowly ionized iron in the circumstellar material that
also causes most of the line of sight absorption. In the HETG spectrum, 
we have found evidence for a weak (equivalent width $\sim$ 30 eV) emission
line at 6.95 keV. This line is identified as Ly$\alpha$ emission line from
hydrogen-like iron and the spectrum does not show emission lines from
helium-like iron. We discuss various emission regions for the
hydrogen-like iron emission line, like gas diffused into the Alfven sphere or
an accretion curtain flowing from the inner accretion disk to the magnetic
poles.

\end{abstract}



\section{Introduction}

GX~1+4 belongs to a rare class of accreting X-ray pulsars with low mass
companion stars. This pulsar, with a spin period of about two minutes,
had a very interesting pulse period history since its discovery (Lewin,
Ricker, \& McClintock 1971) and shows unusual torque-luminosity relation
(Paul, Rao \& Singh 1997, Chakrabarty et al. 1997). Spin downs observed
in GX~1+4 in spite of a large accretion rate corresponding to a persistent
X-ray luminosity of a few times 10$^{37}$ erg s$^{-1}$, suggests a very
high magnetic field strength of the neutron star, 10$^{13-14}$ G (Makishima
et al. 1988) which is yet to be confirmed from spectroscopic measurements.
In medium and low intensity state, GX~1+4 pulse profile has a narrow dip
that goes down to 10--20\% of the average count rate and is believed to
be an eclipse caused by the accretion column (Dotani et al. 1989, Galloway
et al. 2001). The source also shows some low flux episodes associated
with remarkable spectral changes (Naik, Paul \& Callanan 2005).

A low mass companion star and possibility of a very wide orbit with orbital
period of about 300 days (Pereira, Braga \& Jablonski 1999) or more
(Hinkle et al. 2003) also makes GX~1+4
somewhat unique among the X-ray binaries. The only other known symbiotic
star with a neutron star as the secondary is 4U 1700+24 (Galloway et al.
2002). A velocity of about 250 km s$^{-1}$ of the stellar wind (Chakrabarty
\& Roche 1997), low compared to the wind velocity in HMXBs, and a low orbital
velocity of the neutron star results in a large wind accretion radius around
the neutron star. It is also possible that mass accretion takes place via
Roche robe overflow of the red giant companion (Pereira et al. 1999). 
Except for the period when
it was undetected with EXOSAT, GX~1+4 has been one of the brightest hard
X-ray sources in the galaxy. The X-ray spectrum of GX~1+4 is a simple
power-law type, modified at low energies by absorption and at high
energies as an exponential cut-off. A narrow iron K$_\alpha$ line of
varying intensity always adorns the spectrum. Equivalent width of the
iron fluorescence line is found to be related to the column density of
the absorbing material. Based on this relation, Kotani et al. (1999)
suggested that the circumstellar material that causes the absorption
and line emission is in the form of a spherical shell at a distance
of about 10$^{12-13}$ cm from the neutron star.

We have investigated the X-ray spectrum of GX~1+4 at high energy resolution
using an observation with the High Energy Transmission Grating (HETG)
of the Chandra X-ray Observatory. This data was supplemented with a
simultaneous observation with the Rossi X-ray Timing Explorer (RXTE). The
photoionized wind characteristics of GX~1+4 is expected to be different from
the well known high mass X-ray binary systems like Cen X-3 (Wojdowsky et al.
2003), Vela X-1 (Sako et al. 1999, Schulz et al. 2002) and GX~301--2
(Watanabe et al. 2003) due to differences in wind density and velocity.
Here we present the observations and the data reduction method in \S2, timing
and spectral analysis in \S3 followed by a discussion on the results
obtained in \S4.

\section{Observations and data reduction}
 
GX~1+4 was observed with Chandra high energy transmission grating
and the advanced CCD imaging spectrometer (Weisskopf et al. 2002)
in 2002 from August 5, 21:34 UT to August 6 14:58 UT, for a duration
of 57 ks. The HETG consists of
two transmission gratings, a Medium Energy Grating (MEG) with an energy
range of 0.4--5.0 keV and a High Energy grating (HEG) with an energy range
of 0.8--10.0 keV. The dispersed grating spectra are imaged by
the imaging spectrometer. Based on the spectral information from the CCDs,
the grating spectra can be sorted for various grating orders, giving a
low background and a first order spectral resolution of 0.023 \AA\ for
the MEG and 0.12 \AA\ for the HEG. For Chandra HETG observation
with ACIS-S, event1 mode data was reprocessed using the package Chandra
Interactive Analysis of Observations (CIAO, version 3.0) following the CXC
guidelines.  The zeroth-order CCD spectrum was highly affected by pile-up
resulting in a low count rate and hard spectrum. We have, therefore,
not used the zeroth-order CCD spectrum for any further analysis.
For extraction of the grating spectrum, the zeroth-order source
position was determined by analyzing the HEG/MEG dispersion lines and
the data read-out streak of the zeroth order image.

For part of the Chandra observation, GX~1+4 was also observed
simultaneously with the pointed instruments of the RXTE giving a wider
spectral coverage. For spectral analysis, we have used data from the 
Proportional Counter Array (PCA) of the RXTE (Jahoda et al. 1996) for
a duration of 13 ks within the time range of 06:07--13:25 UT on
August 6, 2002. FTOOLS version 5.2 was used for reduction of the RXTE data and
timing analysis. For timing analysis, light curve from a longer time 
base between August 5 23:38 UT to August 6 14:34 UT was used with an 
useful exposure of 28 ks. The PCA standard 1 mode data were used for 
timing analysis, which covers the entire spectral band of the detectors
i.e, 2-60 keV. Standard 2 mode data in the band of 3-30 keV from three
proportional counter units were used for spectral analysis.

\section{Data analysis}

\subsection{Timing analysis} 

From the long term light curve obtained with the RXTE All Sky Monitor it
appears that during the present observation and for about 50 days prior
to this, the source was in a somewhat low intensity state. The 2--20 keV
flux during this observation was 3.8$\times$10$^{-10}$ erg cm$^{-2}$ s$^{-1}$
(see the section on spectral analysis).

The standard 1 mode data of RXTE-PCA has a time resolution of 0.125
s, and the Chandra HETG observation was made with a time resolution
of 3.24 s. Lightcurves from both RXTE and Chandra were barycenter
corrected for timing analysis. Pulses were clearly detected both with
Chandra and RXTE with a pulse period of 138.170 $\pm$ 0.001 s. The 
lightcurves are shown in Figure 1 with a bin size equal to the pulse period.
Strong pulse to pulse intensity variations can be seen clearly, even in the
Chandra data. The pulse profiles, as shown in Figure 2, have a narrow dip,
characteristic of GX~1+4 in medium and low state. The broad characteristics
of the pulse profile measured with Chandra and RXTE seem to be identical
except for some smearing of the Chandra profile due to a poorer time
resolution. Near the two edges of the dip, there is a slight difference
between the two pulse profiles.

\subsection{Spectral analysis}

For spectral analysis of the Chandra grating spectra,  we created
the ancillary response files for the MEG and HEG +ve and --ve first
order spectra using a CIAO script.
The background subtracted positive and negative
first order spectra of both MEG and HEG were combined along with
their response matrices and rebinned to improve statistics for spectral
fitting with a CIAO thread of tools.
At the observed count rate, the MEG and HEG spectrum do not suffer
from any pile-up problem.
Standard procedure prescribed by the RXTE guests observer facility
was adopted for background estimation and spectral response generation.

For spectral analysis we have used the package XSPEC (Arnaud 1996).
We have first fitted the GX~1+4 spectra taken with RXTE-PCA and
Chandra HETG (both HEG and MEG) simultaneously with a model
consisting of a high-energy exponential cutoff power-law, a narrow
emission line at 6.4 keV, and line of sight photoelectric absorption
(Morrison and McCammon 1983, model {\it wabs} in XSPEC).
The galactic ridge emission and CXB background components were
separately included in the model for RXTE-PCA spectrum with
appropriate normalizations (Valinia \& Marshall 1998).
For the RXTE-PCA spectrum, an absorption edge at fixed energy
of 4.8 keV was also included in the model to account for the
xenon L-shell absorption related uncertainty. The relative
normalization of the model was allowed to vary between the
three instruments and the HEG/MEG normalization constants were
about 90\% of the same for RXTE-PCA. Residuals of the HEG
spectral fit shows some excess photons around 6.9 keV and we
have therefore included another one line component at this
energy. The equivalent hydrogen column density of the absorbing
material was found to be 2.3$\times$10$^{22}$ atoms cm$^{-2}$
and the power-law component has a photon index of 0.77,
cutoff energy at 5.6 keV and e-folding energy of 9.8 keV.
There is indication of a very weak soft excess at energies below
1.5 keV, that can be fitted to a blackbody type spectrum with
a temperature of $\sim$0.1 keV. However, we refrain from adding
such a component to the spectral model as a very weak soft excess can
also be produced by small inhomogeneities in the line of sight
absorption. The Chandra-HETG and RXTE-PCA spectra are shown together
in Figure 3, along with the model components and the residuals to the
best fit model. A $\chi^{2}$ of 623 was obtained for 706 degrees
of freedom ($\chi_\nu^{2}$ = 0.88), which is an acceptable fit.
Parameters of the best fit spectral model are
given in Table 1. The 2--20 keV flux during this observation is
3.8$\times$10$^{-10}$ erg cm$^{-2}$ s$^{-1}$, corresponding to
a source luminosity of 4.2$\times$10$^{36}$ erg s$^{-1}$ at
an assumed distance of 10 kpc. We would like to mention here that
there is a substantial uncertainty in the actual distance to GX~1+4,
which affects the luminosity estimate.

To investigate the line emissions in detail, we further analyzed
the MEG and HEG spectra separately. The MEG spectrum (0.3--5.6 keV)
fits well with an absorbed power-law and there is no prominent
emission line in the MEG energy band. To inspect the iron K shell energy
range in detail we have fitted a power-law model to the 5.5--7.5
keV spectrum from HEG with a line at 6.4 keV. The iron fluorescence
line has a width of 12$\pm$8 eV, smaller than grating energy resolution
of $\sim$35 eV at 6.4 keV. The residual to this model clearly shows
some excess around 6.95 keV and inclusion of another line component
at this energy reduced the $\chi^2$ by 7 for 93 degrees of freedom,
corresponding to a F-test false detection probability of
2.0$\times$10$^{-3}$. The center of the second emission line has
an energy of 6.951$\pm$0.025 keV, the uncertainty
quoted here is for 90\% confidence limit. The second emission line has
a nominal width of 13 eV, but with limited photon statistics, 90\%
confidence limits on the line width is 0-40 eV. The HEG spectrum
around the iron K shell energy band along with the best fit model
is shown in Figure 4. We unambiguously
identify this line as Ly$\alpha$ emission from hydrogen-like
iron. The equivalent widths for the
6.40 and 6.95 keV lines are 71$^{+33}_{-16}$ eV and 32$^{+31}_{-20}$
eV respectively. The errors quoted here are for 90\% confidence level.
The HEG spectrum
does not show presence of any line from helium-like iron. We
have determined an equivalent width upper limit of 17 eV
for any emission line in the energy range of 6.7$\pm$0.1 keV with
90\% confidence level.

\clearpage

\begin{deluxetable}{llll}

\tablecaption{Spectral Parameters of GX~1+4}

\tablenum{1}

\tablehead{ \colhead{Parameter} & \colhead {Unit} & \colhead {Value} & \\}

\startdata

Column density ({N$_{\rm H}$}) & {(10$^{22}$ cm$^{-2}$)} & 2.30 $\pm$ 0.04 & \\
Photon index ($\Gamma$)& -- & 0.768 $\pm$  0.005 & \\
Cut-off energy (E$_{\rm C}$) & (keV) & 5.65 $\pm$ 0.06 & \\
E-folding energy (E$_{\rm F}$) & (keV) & 9.82 $\pm$ 0.08 \\
Line centre & (keV) & 6.400 $\pm$ 0.005 & \\
Line flux & (10$^{-4}$ ph cm$^{-2}$s$^{-1}$)&1.7$\pm$0.4\\
Line width & (eV) & 12$\pm$8&\\
Eq. width & (eV) & 71$^{+33}_{-16}$ & \\
Line centre & (keV) & 6.951 $\pm$ 0.025& \\
Line width & (eV) & $<$42&\\
Eq. width & (eV) & 32$^{+31}_{-20}$ \\ 
Line flux & (10$^{-4}$ ph cm$^{-2}$s$^{-1}$) &0.7$^{+0.7}_{-0.4}$ &\\
$\chi_\nu^{2}$ (DOF)& --& 0.88 (706)&\\

\enddata

\tablecomments{Spectral parameters of GX~1+4 measured from a simultaneous
fitting of the Chandra HETG and RXTE-PCA spectra. The iron line parameters 
are measured using HEG data only. The errors given here are for 90\% confidence
limits.}

\end{deluxetable}

\clearpage

\section{Discussion: Origin of the two iron K shell lines}

The 6.4 keV K$_\alpha$ fluorescence line from neutral or lowly ionized iron has
been observed in GX~1+4 many times in the past. The line equivalent width
is usually found to have a dependence on the absorption column density,
implying that the line originates from a spherical circumstellar shell.
The very low ionization state of the iron, as derived from the ratio of
the line centre energy to the K shell absorption edge energy indicates
that the circumstellar shell is at a large distance of about 10$^{12}$ cm
from the neutron star (Kotani et al. 1999). In the present observation,
the equivalent absorption column density, determined very accurately with
the Chandra detectors, is rather low, 2.3$\times$10$^{22}$ atoms cm$^{-2}$.
We note here that the total Galactic HI column density in the direction of
GX~1+4, measured from the HI map (Dickey \& Lockman, 1990) is 0.3$\times$10$^{22}$
atoms cm$^{-2}$, a
small fraction of the equivalent hydrogen absorption column density measured
from the X-ray spectrum. Assuming a solar abundance of iron, a uniform
spherical shell with such a low column density is expected to give a
neutral fluorescence line equivalent width of only about 25 eV. A larger
iron abundance compared to the solar value or a non-spherical distribution
of the circumstellar material can also produce a different line equivalent
width. It is also possible that a part of the observed equivalent width of
70 eV may originate from re-processing of hard X-rays in the accretion disk,
like the other low mass binary pulsar Her X-1 (Endo et al. 2000). If the
orbital period of GX~1+4 is correct, the companion star subtends a substantial
solid angle to the compact object (Pereira et al. 1999, also see Hinkle et al
2003), hence surface of
the companion star can also be a X-ray reprocessing region producing part
of the iron fluorescence line. Wind velocity of the red giant companion star
(250 km s$^{-1}$), or rotational velocity of matter in the outer accretion disk,
both can produce an intrinsic line width of a few eV, comparable to the present
measurement.

One important discovery from the present work is the presence of a
Ly$\alpha$ line from hydrogen-like iron and the conspicuous absence
of any emission line from helium-like iron. In the X-ray irradiated
wind of GX~1+4, we do not detect any emission line from helium-like
iron, which in some other sources like in Vela X-1 (Sako et al. 1999,
Schulz et cl. 2002) and Cen X-3 (Wojdowsky et al 2003) indicates
presence of onion-like structures with continuously changing ionization
degree and different ionization level of metals. The uniqueness of GX~1+4
is a large Alfven radius and a comparatively low wind velocity of the companion
star. It appears that in GX~1+4, the 6.4 keV and the 6.95 keV lines originate
from two completely different regions. One possibility is that while the 6.4
keV fluorescence line originates from cold material very far from the
neutron star (circumstellar shell, and/or accretion disk, surface of
companion star), the hydrogen-like line originates from a hot gas inside
or near the Alfven shell, where the ionisation parameter is likely to be high.
Similar X-ray spectrum, with a hydrogen-like iron emission line and absence
of helium-like line has been detected in Her X-1 (Jimenez-Garate et al. 2005),
another one accreting X-ray pulsar with a medium mass companion star.
If the 6.95 keV line is produced from plasma very close to the neutron star
(a few $r_g$), gravitational redshift will cause line centre energy to shift
to lower value. At the same time, the differential redshift over the size of
the plasma will also cause the line to broaden. Since the line detected with
HETG has an upper limit of 40 eV on its width (90\% confidence limit), we do
not think that this line is produced very close to the NS surface where
gravitational redshift would be important.

\subsection{Line emission from gas inside the Alfven radius}

We first explore whether the hydrogen-like iron line observed in the GX~1+4
spectrum with Chandra-HETG can be produced from a mixture of hydrogen-like
and fully stripped iron inside the Alfven shell. A fraction of
the accretion material, while being transported from the inner accretion
disk to the accretion column via the magnetospheric surface, may diffuse
inside the magnetospheric boundary. We presume that outside the
magnetosphere, the wind density is very low except in the disks. The iron
atoms in the outer accretion disk or in the circumstellar material
are near neutral while iron inside the Alfven radius are hydrogen-like
or fully stripped.
Assuming a neutron star radius of 10 km, and a mass of 1.4 M$_\sun$,
the Alfven radius of GX~1+4 in the present state of luminosity 
(4.3$\times$10$^{36}$ erg s$^{-1}$) can be expressed as (Frank, King \&
Raine 2002)
$$
r_{\rm M} = 2.9 \times 10^{8}~{\left({\rm M}\over{{\rm M}_\sun}\right)}^{1\over7}~{\rm R}_{6}^{10\over7}~{\rm L}_{37}^{-{2\over7}}~{\rm B}_{12}^{4\over7}~{\rm cm}
= 3.9 \times 10^{8}~{\rm B}_{12}^{4\over7}~{\rm cm}
$$
where
B$_{12}$ is the surface magnetic field strength in unit of 10$^{12}$ Gauss.
We mention that the discussion here and in the next section is subject to
the luminosity based on an assumed distance of 10 kpc, which is rather
uncertain. It is also subject to a bolometric correction to the luminosity
compared to the measured flux in the 2-20 keV band, which we cannot do
in absence of simultaneous hard X-ray measurement.

For some fraction of the iron atoms to remain in hydrogen-like state
and at the same time not to have a significant amount of helium-like 
iron (since we do not detect any helium-like line), the ionization
parameter ($\xi$ = ${{\rm L}\over{nr^{2}}}$) at the Alfven radius
should be in the range of 3.5 $<$log($\xi$)$<$ 4.5, where $n$ is the particle
density and $r$ is the distance from the neutron star. For log($\xi$) = 3.5,
the particle density at the Alfven radius should be 
$$n = 8.9 \times 10^{15}~{\rm B}_{12}^{-{8\over7}}~cm^{-3}$$
with a corresponding Thompson optical depth of 
$$\tau = 2.3 \times {\Delta r\over r}~{\rm B}_{12}^{-{4\over7}}$$
where $\Delta r$ is the thickness of the diffused gas inside the
Alfven radius. 

We have simulated the line emissivity from such a photoionized spherical
gas cloud with an input power-law X-ray spectrum at the centre, as measured
during the present observation. In the simulation, carried out with the XSTAR
package\\
(http://heasarc.gsfc.nasa.gov/docs/software/xstar/xstar.html),
different values of neutron star surface magnetic
field was assumed (B = 10$^{12}$, 10$^{13}$, and 10$^{14}$ Gauss), and
different values for the uniform
gas density was chosen in such a way
that the ionization parameter in each case has a value of
log$\xi$ = 3.5, 4.0, or 4.5 at the outer boundary of the sphere.
The simulated 6.95 keV line flux from this ionized gas cloud was found to
be comparable to the observed line flux of 7$\times$10$^{-5}$ photons
cm$^{-2}$ s$^{-1}$ for the case where B = 10$^{14}$ Gauss, log($\xi$) = 3.7,
and $\tau = 0.1$. A smaller magnetic field (i.e. smaller Alfven radius)
and larger gas density can also produce an equal amount of line flux with
a larger Thompson optical depth.

However, such a scenario seems to be implausible because the mass required
to be present inside the Alfven radius is very large, of the order of 
$10^{19}$ gm. Even with a slow infall onto the neutron star at 10\% 
of the free fall velocity, this should produce an unpulsed X-ray luminosity
comparable to or greater than the total X-ray luminosity measured during 
this observation. In addition, the diffusion rate through a strong magnetic
field is expected to be very low (Elsner and Lamb, 1984) compared to the
mass required for production of the observed line intensity.

\subsection{Line emission from gas flowing onto the neutron star along
the magnetic field lines}

We subsequently examined the line flux expected from material flowing
onto the neutron star in the shape of an accretion curtain (Miller 1996)
over the dipole magnetic field structure as it leaves the inner accretion
disk. For a crude estimate of mass in this shell type structure, the material
required to produce the observed X-ray luminosity is assumed to flow at
10\% of the free-fall time-scale and the solid angle subtended by the
accretion curtain to the neutron star is assumed to be $\pi$ steradian.
From simulations with XSTAR we found that irradiation of the accretion
curtain by the central X-ray source can produce the observed flux of
the 6.95 keV line without producing any other prominent emission line
if the thickness of the accretion curtain is about 10\% of the Alfven 
radius. This is in accordance with the fact that in the accretion disks
surrounding strongly magnetised neutron stars, the transition zone is
known to be rather broad (Ghosh and Lamb, 1979). If the boundary layer
is sharp, it causes a larger material density in the accretion curtain
and low ionisation parameter and therefore cannot produce the observed
hydrogen-like iron line, unless the Alfven radius is too small.

Therefore, it appears that the observed line emission from hydrogen-like
iron is produced by X-ray irradiation of gas flowing from the inner accretion
disk onto the neutron star along the magnetic field lines rather than
material diffused into the Alfven sphere. The 6.95 keV line flux depends
critically on the ionisation parameter, and hence on the mass accretion rate
and Alfven radius / magnetic field strength of the neutron star. This
may explain why such a line has not been detected in the HETG spectrum
of other accretion powered X-ray pulsars, except in Her X-1 (Jimenez-Garate
et al. 2005). In GX~1+4, this hydrogen like iron emission line may have a
significantly different equivalent width at a different mass accretion rate.
We also note that the detection of this line with the present Chandra
HETG observation is only at 3$\sigma$ level and the upper limit for 
emission lines from helium-like iron is not very small compared to the
line flux of the hydrogen-like iron. If the 6.95 keV line does indeed
originate from the accretion curtain, a flux variation of this line with
pulse phase
would be expected while the 6.4 keV line flux is independent of the pulse
phase (Paul \& Nagase, 1999). We have carried out a pulse phase resolved
spectroscopy of the HEG data with four phase bins, but the very low flux of
the 6.95 keV line caused a very low sensitivity for flux variation measurement
and the result was inconclusive. The future ASTRO-E2 X-ray calorimeter,
which will have a much higher sensitivity for narrow emission lines and
a factor of five better energy resolution in this energy band will be very
valuable to unravel the actual line emission mechanism in GX~1+4.

\begin{acknowledgements}

The authors like to thank an anonymous referee for suggestions that
helped to improve the paper. We thank Chandra X-ray Observatory for
providing the archival data and the data reduction software. BP thanks
Randall Smith for discussions about Chandra data reduction.

\end{acknowledgements}

\clearpage

\begin{figure}
\begin{center}
\includegraphics[totalheight=3.5in,angle=-90]{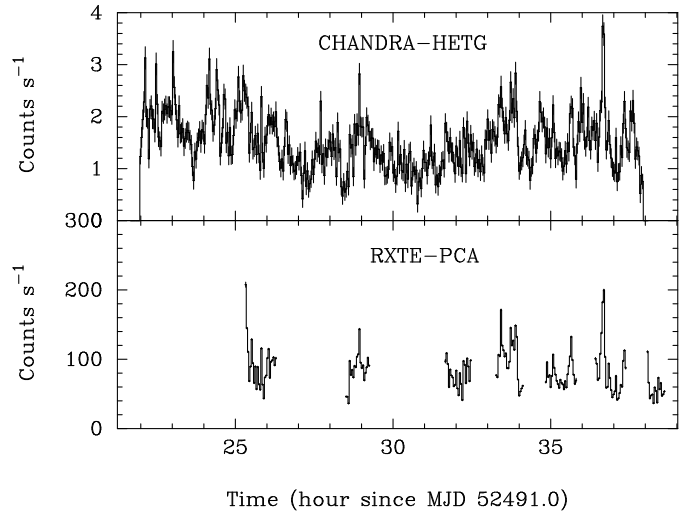}
\end{center}
\caption{
Background subtracted simultaneous lightcurves of the Chandra-HETG in the
0.5--8.0 keV band and RXTE-PCA in the 2.5--60.0 keV band are shown here. The
PCA light curve is summed for three detectors and shown only when three
detectors were active. The plot is made with a binsize of 138.17 s,
same as the spin period during this observation.}
\end{figure}

\begin{figure}
\begin{center}
\includegraphics[totalheight=3.0in,angle=-90]{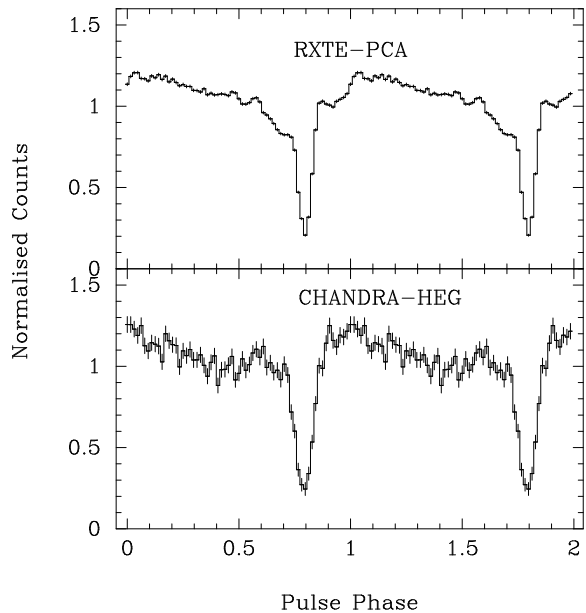}
\end{center}
\caption{
The background subtracted pulse profiles of GX~1+4 obtained with the RXTE-PCA
and Chandra-HETG for the same energy band as in Figure 1. are shown here for 
two cycles. The epoch for phase zero of the pulse profile is taken arbitrarily.}
\end{figure}

\begin{figure}
\begin{center}
\includegraphics[totalheight=4in,angle=270]{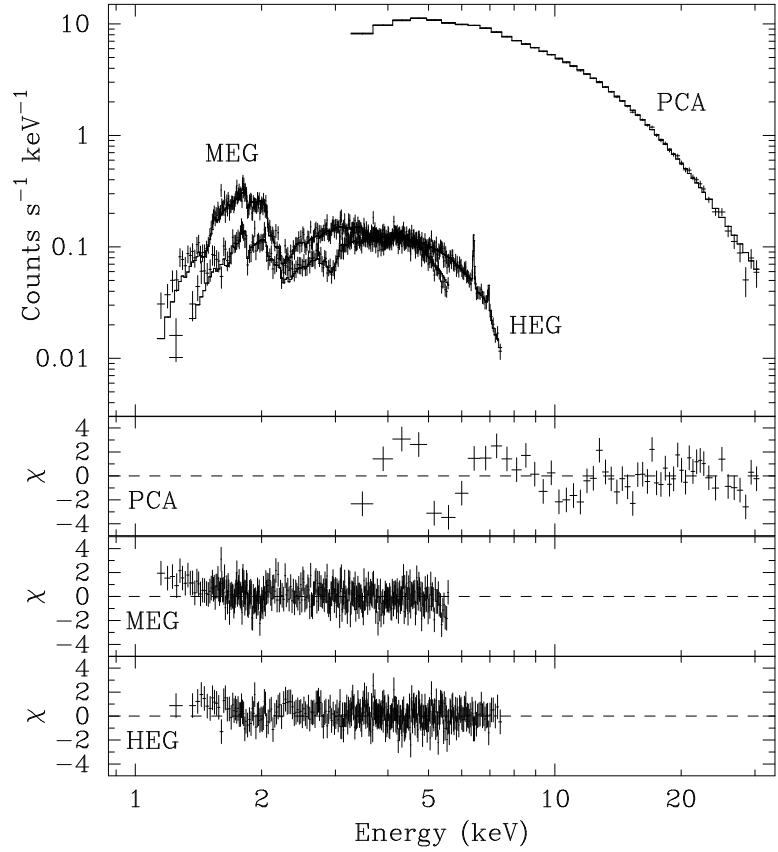}
\end{center}
\caption{The observed X-ray spectrum of GX~1+4 obtained
with RXTE-PCA and Chandra HETG (HEG and MEG) are shown here
along with the best fitted model folded with the detector and
telescope response. The spectra were fitted simultaneously with
a high-energy exponential cutoff power-law with line of sight
absorption and two Gaussian emission lines. Lower panel shows
the residuals to the best fitted model.}
\end{figure}

\begin{figure}
\begin{center}
\includegraphics[totalheight=4in,angle=-90]{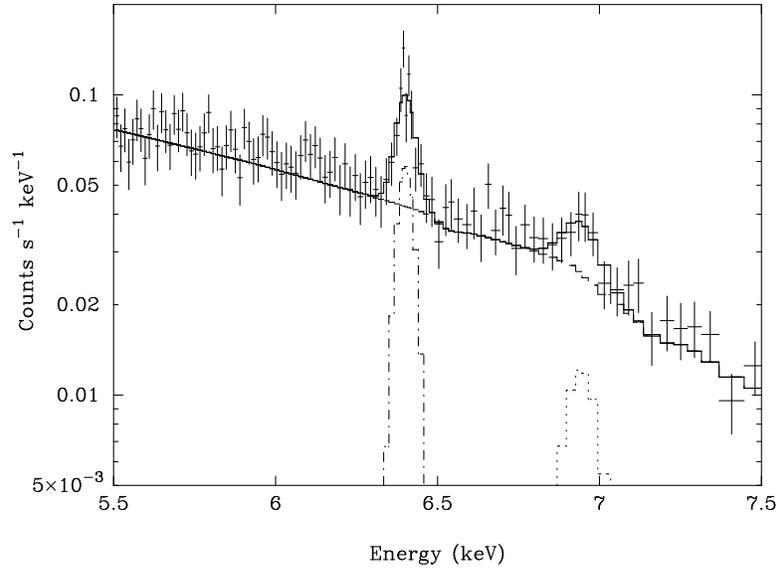}
\end{center}
\caption{
The HEG spectrum of GX~1+4 fitted in the 5.5--7.5 keV range
with a power-law and two Gaussian lines. In this plot the
bin size is larger at higher energy, $\sim$16 eV at 6.4 keV
and $\sim$25 eV at 6.95 keV. The HEG instrument resolution is
about 35 eV at 6.4 keV.}
\end{figure}

\end{document}